\documentclass[%
 reprint,
 amsmath,amssymb,
 aps,
]{revtex4-2}
\usepackage{mathbbol}
\usepackage{graphicx,color}
\usepackage{amsmath,amssymb,bm,amsthm,amstext}
\usepackage{mathtools}
\usepackage{simplewick}
\usepackage{braket}
\usepackage{qcircuit}
\usepackage{array}
\usepackage{soul}
\usepackage{dcolumn}
\begin{document}

\pagecolor{white}

\preprint{APS/123-QED}

\title{Nuclear Induction Lineshape Modeling via Hybrid SDE and MD Approach}
\thanks{Correspondence: lsbouchard@ucla.edu (L.-S. B.)}%

\author{Mohamad Niknam}
\affiliation{Department of Chemistry and Biochemistry, University of California Los Angeles, 607 Charles E. Young Drive East, Los Angeles, CA 90095-1059, USA}
\affiliation{Center for Quantum Science and Engineering, UCLA}
\author{Louis-S. Bouchard*}
\affiliation{Department of Chemistry and Biochemistry, University of California Los Angeles, 607 Charles E. Young Drive East, Los Angeles, CA 90095-1059, USA}
\affiliation{Center for Quantum Science and Engineering, UCLA}
\email{lsbouchard@ucla.edu}%

\date{\today}

\begin{abstract}
The temperature dependence of the nuclear free induction decay in the presence of a magnetic-field gradient was found to exhibit motional narrowing in gases upon heating, a behavior that is opposite to that observed in liquids. This has led to the revision of the theoretical framework to include a more detailed description of particle trajectories, since decoherence mechanisms depend on histories. In the case of free diffusion and single component, the new model yields the correct temperature trends.  Inclusion of boundaries in the current formalism is not straightforward. We present a hybrid SDE-MD (stochastic differential equation - molecular dynamics) approach whereby MD is used to compute an effective viscosity and the latter is fed to the SDE to predict the lineshape. The theory is in  agreement with experiments.  This two-scale approach, which bridges the gap between short (molecular collisions) and long (nuclear induction) timescales, paves the way for the modeling of complex environments with boundaries, mixtures of chemical species and intermolecular potentials.
\end{abstract}

\maketitle


\section{\label{sec:Intro}Introduction}

During the mid 20th century, when the first nuclear magnetic resonance (NMR) experiments were performed in liquids, magnetic-field inhomogeneity was the limiting factor that determined the properties of nuclear induction.  Rapid signal decay is problematic on so many levels, especially because it limits the  spectral resolution.  This led to the development of the Hahn echo~\cite{HahnEcho}, where the effects of external magnetic field inhomogeneity are removed by a $\pi$-rotation about the $x$ axis in a time-reversal experiment. Spin echoes enabled more accurate studies of  intrinsic spin-spin relaxation mechanisms and  molecular structures. Years later, the deliberate creation and modulation of magnetization by magnetic-field gradients formed the basis of modern techniques in magnetic resonance imaging (MRI)~\cite{mansfield1977multi,lauterbur1973image}.

Given the ubiquitous nature and uses of magnetic-field gradients in modern magnetic resonance experiments, it is imperative that  the nuclear response function and its dependence on the sample under study in the presence of applied gradients be properly understood.  A correct understanding of the true limits and capabilities of the experiment will enable accurate interpretation of experimental results, as well as inform future developments and applications.   In organic and biological chemistries, for example, NMR spectroscopy is routinely used as a tool to analyze solution content and composition for chemical species' identities and relative content.  Limiting factors that compromise linewidth must be minimized or removed so that spectra can be obtained of sufficiently high resolution to elucidate molecular structures.   The diffusion of molecules can cause undesirable effects such as altering lineshape or be used advantageously, for example, by helping probe transport phenomena.  While diffusion can sometimes lead to sharper lines (e.g., increased spectral resolution due to diffusional averaging of the intermolecular magnetic dipole interaction), the presence of residual gradients does create a complex relationship between nuclear induction and sample properties and geometry that obscures the interpretation of results that rely on lineshape analysis. Diffusion-weighted readouts have enabled imaging of gases in lungs~\cite{diffMRIlung}, {\it operando} monitoring of chemical reaction~\cite{lysova2010,gladden99,nature13} as well as biophysical or mechanical properties of gels and biological tissues~\cite{van1996unraveling,qin2013combining}. In materials science and engineering, diffusion processes are essential to fabrication and synthesis. 

In light of these various effects, the observation that in gases lines become narrow with temperature was rather surprising~\cite{nature13,PRL15,JCP18} given that the conventional theory predicted a broadening instead~\cite{HahnEcho,HerzogHahn,PFITSCH1999,slichterbook}.   In liquids, a broadening is observed experimentally, and this is in agreement with  theory.  This led to a revision of the lineshape theory in gases accounting for the time history of rapid molecular motions below the timescale of the NMR experiment~\cite{Tongcang13}.  It is critical to account for the history of molecular motions when the process involves motional averaging due to collisions.  Conventional theory ignores the histories of molecular diffusion and instead assumes a Gaussian probability distribution for the accumulated phase of the spins. The Gaussian assumption on phase accumulation, which neglects the details of molecular trajectories, only appears justified in liquids, where many collisions occur on the NMR timescale. But in gases
the average separation between molecules is much larger and this assumption is more difficult to justify.  A drawback of the new theory is its increased complexity.  While it provides good agreement between theory and experiments in the case of free diffusion, it is unclear how one should model the effects of realistic boundaries.  Stochastic differential equations (SDE) can be used to model molecular trajectories but this still requires validation against experiments.

In this study, we bridge the gap using molecular dynamics (MD) simulations and coupling it to an SDE in order to describe the nuclear induction. By modeling viscosity of fluids using MD, the lineshape is then derived from the viscosity via its connection to the SDE. Our simulations of particle collisions and trajectories using Lennard-Jones interaction reveals opposite trends in liquid and gas particle viscosity coefficients as function of temperature. This in turn can be used to describe the lineshape trends in gases and liquids. This work establishes feasibility of the hybrid SDE-MD approach and paves the way for modeling of complex interactions with boundaries and other components. The advantages of modeling viscosity by MD include the possibility of including realistic boundaries, multiple components or varying the details of intermolecular interactions. Indeed, models of effective viscosity have already been developed for such situations~\cite{Starov01,Breugem07,Rudyak18,RizkPRL2022}.

\subsection{Review of Conventional Theory}

Assume that an ensemble of spins is placed in an external magnetic field and a coherent superposition between the states $\ket{\uparrow}$ and $\ket{\downarrow}$ is created. In an inhomogeneous field, the phase accumulation for each spin is proportional to the local field experienced by the spin over time.  Phase accrued by nuclear spins can be refocused by application of a $\pi$ pulse which inverts the direction of spin rotations, or simulating a ``time-reversal'', forming a Hahn echo~\cite{HahnEcho,CP54,Torrey56,CallaghanBook91,Stejskal65}. If the spins are fixed (frozen) in space, evolution is described with a unitary propagator and the entropy remains constant. With diffusion (e.g. liquids and gases) not all initial phases within a spin ensemble can be recovered and the Hahn echo signal will be smaller. 
Assuming that the molecules undergo a random walk, it has been postulated that they sample random phase increments from a Gaussian distribution~\cite{HahnEcho}. In the presence of a magnetic-field gradient $\mathbf{g}$ the phase accrued by a spin from time $0$ to $t$ is:
\begin{equation}
\label{eq:phase}
\Delta \phi(t) = \int_0^t \omega(t') dt' = \gamma_n \int_0^t \mathbf{g \cdot r}(t') dt',
\end{equation}
where  $\omega(t)$ is the time-dependent frequency of a moving spin, $\gamma_n$ is the nuclear gyromagnetic ratio, and $\mathbf{r}(t)$ is the time-dependent position of the diffusing particle.
The nuclear induction signal is weighted by the ensemble average of these phase factors, written with probability distribution  $P(\Delta \phi)$ 
\begin{equation}
\left<\exp(i \Delta \phi(t))\right>=\int_{-\infty}^{\infty} P(\Delta\phi)\exp(i \Delta \phi)d(\Delta \phi).
\end{equation}
To alleviate the notation we will drop $t$ from the notation and write $\Delta \phi(t)=\Delta \phi$. Assuming a Gaussian distribution for $P(\Delta\phi)$ with variance $\langle (\Delta \phi)^2\rangle$ proportional to $t$, we get
$$ S(t)= \left<\exp(i \Delta \phi )\right>=\exp(-\left<(\Delta\phi)^2\right>/2). $$
Thus for the free diffusion under steady magnetic field of a linear gradient $\mathbf{g}$ and duration $t$, attenuation of the NMR signal $S(t)$ was found to be
\begin{equation}
\label{eq:sigt3}
S(t)=\exp{[(-1/3)\gamma_n^2g^2Dt^3]},
\end{equation}
where $D$ is the self-diffusion coefficient and the $t^3$ decay arises from self-diffusion in presence of a field gradient~\cite{Stejskal65,CallaghanBook11}.  
This equation shows that for unrestricted diffusion, as $D$ gets larger a wider range of $\Delta\phi$ is sampled resulting in faster damping of the signal.  In the Hahn echo experiment, $90^{\circ}-\tau-180^{\circ} - \tau$, at $t=2\tau$ the decay function is similar:
\begin{equation}
\label{eq:han1}
S(2\tau)=\exp(-\frac{2}{3}\gamma_n^2 g^2 D \tau^3).
\end{equation}
The $t^3$ dependence has been validated experimentally for liquids. This form of signal can be used to describe a Carr-Purcell-Meiboom-Gill (CPMG) experiment with $n$ refocusing $\pi$ pulse and interpulse delay of $2\tau$~\cite{CP54,MG58}. The signal at $t=2n\tau$ decays according to
\begin{equation}
\label{eq:han2}
S(2n\tau)=\exp\left(-\frac{2}{3}\gamma_n^2 g^2 D \tau^2 (n\tau) \right).  
\end{equation}
As a result, the signal attenuation at time $t=2n\tau$ can be reduced by decreasing time delay between the pulses $\tau$ while increasing number of pulses $n$.  Equations~(\ref{eq:han1}) and~(\ref{eq:han2}) have been extensively validated in experiments for the case of liquids.  The assumption of Gaussian-distributed phase increments is reasonable for liquids because of the short mean-free paths and large number of collisions during time intervals of duration $\tau$.

\subsection{Problems With Conventional Theory}

In the case of gas molecules, which are characterized by longer mean free paths, the assumption of randomly distributed phase increments becomes more difficult to justify~\cite{PRL15,PRLreb}.  A simple example can be used to illustrate the need for a more sophisticated model.  Take the Ornstein-Uhlenbeck process, which is used to model Brownian particle diffusion.  The velocity process obeys the SDE:
$$ dv = - \gamma v dt + \frac{\Gamma_f}{M} dW(t), $$
where $\gamma^{-1}$ is a damping time constant, $dW(t)$ is the increment of the Wiener process $W(t)$ at time $t$ ($dW(t):=W(t)-W(t-dt)$). The strength of fluctuations is $\Gamma_f = \sqrt{2 \gamma M k_B T}$ (fluctuation-dissipation theorem), where $M$ is the mass of diffusing particle and $k_B$ is the Boltzmann constant. After integration
$$ v(t) = \langle v(t) \rangle  + \sqrt{  \frac{2\gamma k_B T}{M}} e^{-\gamma t} \int_0^t e^{\gamma t'} dW(t'), $$
where $\langle v(t) \rangle = \langle v_0 \rangle e^{-\gamma t}$. 
The position process is the time integral of the velocity process, $dx(t) = v(t) dt$, whose solution is (setting $x(0)=0$):
$$ x(t) = \frac{v_0}{\gamma} (1-e^{-\gamma t}) + \sqrt{ \frac{ 2\gamma k_B T}{ M} } \int_0^t dt' e^{-\gamma t'} \int_0^{t'} dW(t'') e^{\gamma t''}. $$
Defining a thermal velocity $v_T = \sqrt{k_B T/ M}$, the distribution of $x(t)$ is (again, assuming $x(0)=0$), 
$$ p(x,t) = \frac{1}{\sqrt{ 2\pi \sigma_x^2(t)}} \exp \left[ - \frac{ (x - v_T(1-e^{-\gamma t})/\gamma )^2}{ 2 \sigma_x^2(t) } \right] $$
where
$$ \sigma_x^2(t):=\frac{ 2 v_T^2}{ \gamma} t - \frac{ v_T^2 }{ \gamma^2 } ( 3 - 4 e^{-\gamma t} + e^{-2\gamma t} ) $$
at short $t$  is not the usual Gaussian with variance $t$ characteristic of long-time particle diffusion (i.e. the Einstein-Fick limit).  
Only at long times ($t \gg \gamma^{-1}$) does the distribution become Gaussian with variance $t$:
$$ p(x,t) \rightarrow \frac{1}{\sqrt{ 4\pi Dt }} \exp \left[ - \frac{ (x-v_T/\gamma)^2 }{ 4 D t} \right] $$
with diffusion coefficient $D=k_B T/M \gamma$.  To compute the accrued phase, which is a time integral of $x(t)$
(c.f. Equation~\ref{eq:phase}), it appears prudent to model the behavior at short times ($t \lesssim \gamma^{-1}$). 
The timescale $\gamma^{-1}$ of the friction coefficient is associated with molecular collisions. 

It has been shown~\cite{PRL15} that Eq.~(\ref{eq:han1}) or~(\ref{eq:han2}) does not describe the nuclear induction decay in gases. Instead, it was observed that in the CPMG experiment:
\begin{equation}
\left<\exp(i \Delta \phi)\right> =\exp(-\gamma^2 g^2 \kappa(2n\tau)).
\end{equation}
Here the decay constant $\kappa$ is a decreasing function of temperature $T$, damping rate $\gamma$, and particle mass $M$, Eq.~(\ref{eq:kappa}). Notice the different powers of $\tau$ inside the argument of the exponential function ($\tau^3$ vs $\tau$). This difference was first noticed by observing the linewidth dependence on temperature~\cite{nature13} and subsequently investigated more in-depth in Ref.~\cite{PRL15,PRLreb}. The conclusion from these studies is that a more detailed description of diffusion effects is needed.  Obviously, exact solutions to the $N$-body problem are not possible; hence we seek suitable models.

\subsection{Generalized Langevin Equation (GLE)}

The GLE is a well-established model of particle diffusion that accounts for memory effects as the molecules undergo diffusion via collisions. In the gas and liquid phase, nuclear spin degrees of freedom are fairly well isolated from spatial degrees of freedom, at least to first order. Consequently, individual molecular collisions, do not completely depolarize the spins.  Instead, decoherence and depolarization processes, which are second-order processes, take place over much longer periods of time.
Decoherence is described by a characteristic decay time $T_2$. The motional part of the NMR signal thus can be written as the expectation value of the spin phase factors [c.f. Equation~\ref{eq:phase}, with $\omega(t)=\gamma_n g \cdot x(t)$ in 1-D] which depend on their position $x(t)$~\cite{CallaghanBook11,PRL15}.  In Ref.~\cite{PRL15} it is assumed that $x(t)$ is a  Gaussian random process that is stationary in the wide sense.  The expectation value
\begin{equation}
\label{eq:signalt}
S(t)=\left< \exp\left(i\int_0^t \omega(t')dt'  \right) \right>
\end{equation}
takes the form:
$$ \exp\left[ i \gamma_n g \int_0^t \langle x(t') \rangle dt' - \gamma_n^2g^2 \int_0^t \langle x(t') x(0) \rangle(t-t')dt' \right]. $$
The second term determines the lineshape of the signal decay.
For liquids, displacements are small $x(t) \approx x(0)$ and this leads to the Einstein-Fick limit $\langle x(t) x(0) \rangle  \approx \langle x(t) x(t) \rangle = 2Dt$, where $D$ is a diffusion coefficient, and the classic Hahn result~(\ref{eq:sigt3}) is recovered.

We note that there is no easy way to compute the integral over the function $\langle x(t') x(0) \rangle$ from first principles in the general case of arbitrary physical conditions.  Even MD simulations fail because $\langle x(t') x(0) \rangle$ decays on short time scales (tens to hundreds of picoseconds) before reaching a nonzero steady state value, after which the time integral is weighted by the monotonically increasing weight $(t-t')dt'$ over much longer time scales (microseconds to seconds).  The initial decay, and any numerical errors associated with it, is then amplified by orders of magnitude.  Instead, a model is needed that bridges the two widely different timescales by analytically solving the integral and reducing it to a function of transport coefficients that are easier to compute. Such a transformation was first presented in Ref.~\cite{nature13}.  The main steps of its derivation, although well known to the fluid dynamics community, are recapped below for convenience.

In the ``weak collision'' regime (see Ref.~\cite{PRL15}) a GLE with memory kernel  describes particle dynamics
\begin{equation}\label{eq:GLE}
M\Dot{v}+\int_0^t\Gamma(t-t')v(t')dt'=\eta_f(t)
\end{equation}
where $M$ is the mass of diffusing particle, $v(t)=\Dot{x}(t)$ and $\Dot{v}$ are particle velocity and acceleration. $\eta_f(t)$ on the right-hand side, represents a time-dependent stochastic force. The memory kernel $\Gamma(t)$, is convoluted with the particle velocity to describe the friction asserted by the viscous dynamics. 
Projecting both sides of the GLE equation with the inner product $\langle v(0) ,\cdot\rangle$ we get
\begin{equation}
    M \left< v(0)\Dot{v}(t)\right>+\int_0^t\Gamma(t-t')\left< v(0)v(t')\right>dt'=0.
\end{equation}
Fluctuation-dissipation, $\langle \eta_f(0) \eta_f(t) \rangle = k_B T \Gamma(t)$, and equipartition, $\left< v(0)v(0)\right>=k_BT/M$, theorems are invoked.  Here, $\nu(t)$ is the integral of the velocity autocorrelation function, $\int_0^t \left< v(0)v(t')\right>dt'$. Then,
\begin{equation}
M\Dot{\nu}(t)+\int_0^t\Gamma(t-t')\nu(t')dt'=k_B T.
\end{equation}
For the memory kernel one often invokes the Ornstein-Uhlenbeck process to model the delayed response of surrounding fluid
\begin{equation}
\Gamma(t)=(\gamma^2/m)\exp(-\gamma t/m),
\end{equation}
where $\gamma$ is a friction coefficient proportional to the viscosity of the fluid and $m$ represents a mass attributed to solvent particles. The solution using $\zeta_{\mp}= \frac{\gamma}{2m}(1\mp \sqrt{1-4m/M})$ is 
\begin{multline}
\nu(t)=\frac{k_BT}{M} \left(\frac{\gamma}{m\zeta_-\zeta_+}+\frac{1}{\zeta_+-\zeta_-}\left[(1-\frac{\gamma}{m\zeta_+})\exp({-\zeta_+t}) \right. \right. \\
\left. \left. -(1-\frac{\gamma}{m\zeta_-})\exp(-\zeta_-t)\right] \right).    
\end{multline}
For the position auto-correlation function one finds 
\begin{multline}
\left< x(t)x(0)\right>=\frac{k_BT}{M(\zeta_+-\zeta_-)}\left[\zeta_+^{-1}(1-\frac{\gamma}{m\zeta_+})\exp({-\zeta_+t})\right. \\
\left. \zeta_-^{-1}(1-\frac{\gamma}{m\zeta_-})\exp({-\zeta_-t}) \right].
\end{multline}
In the case that $\gamma t/m$ is sufficiently large, the Ornstein-Uhlenbeck kernel rapidly decays. At this point, it is possible to introduce a dependence on viscosity by invoking Stokes' drag law $\gamma = 6 \pi \eta R$ where $R$ is the radius of the ``Brownian particle'' and $\eta$ is the shear viscosity. The validity of Stokes' law is predicated on the assumption that the Brownian particle is much larger than the solvent particles ($M \gg m$).  In the case of self-diffusion all particles are identical.  Hence, the Stokes' law appears unjustified.  However, the proportionality between $\gamma$ and $\eta$ is always correct~\cite{bib:ein3}.  In fact, Einstein developed the concept of ``effective viscosity'', which has been used to describe the effective viscosity of lubricants~\cite{bib:ein1,bib:ein2,bib:ein3,bib:ein4}.

We are now in a position to replace the integral by a function of the viscosity, a transport coefficient that is easily computed from MD simulations with good accuracy.  Viscosity is essentially a coarse-grained quantity describing the relaxation of momentum density to its equilibrium value, after a perturbation. It is directly related to the velocity autocorrelation function. Stokes' law has been extended to the frequency domain to describe dynamical and dissipative effects in rheology~\cite{zwanzig1970hydrodynamic,mason1995optical,cordoba2012elimination}. The relationship between the Fourier representation of frequency dependent viscosity and the frequency-dependent friction coefficient $\tilde{\gamma}(s)$ is
\begin{equation}
\tilde{\eta}(s)=\frac{\tilde{\gamma}(s)}{6 \pi R}
\end{equation}
where $s$ is the complex frequency in Laplace domain and $\tilde{\gamma}(\omega)=\int_0^{\infty}\Gamma(t)\exp(i\omega t)$~\cite{zwanzig1970hydrodynamic}.  This relationship, which has been termed the ``correspondence principle''~\cite{cordoba2012elimination}, provides a direct link between the memory function and viscosity. The memory kernel itself encodes the response of particles at all frequencies to collisions and boundary conditions. Therefore, we expect the viscosity to be sensitive to boundary conditions and intermolecular potentials as well. In particular, we expect that changes in the lineshape of the NMR signal, c.f. Eq.~(\ref{eq:signalt}), as a function of temperature, to be reflected in the memory function as well as viscosity coefficient. 

{\it The case of gases.}  A model for the viscosity and its temperature dependence is needed~\cite{Brush62}. Dynamic viscosity of gases was described by Sutherland as
\begin{equation}\label{eq:gasviscform}
\eta=\frac{\mu_0 (T_0+C)}{T+C}(\frac{T}{T_0})^{3/2}\sim \frac{T^{3/2}}{T+C},
\end{equation}
where $C$ is Sutherland's constant, and $\mu_0$ is the viscosity at temperature $T_0$. At high temperatures viscosity grows with the square root of the temperature, $\eta \sim T^{1/2}$, while at low temperature $\eta \sim T^{3/2}$. 
Using Stokes' drag law, we rewrite the nuclear induction signal as~\cite{PRL15} 
\begin{equation}
S(t)=\exp(-\gamma_n^2g^2\kappa t),
\label{eq:nmrsignal}
\end{equation}
with
\begin{equation}\label{eq:kappa}
\kappa(T)=\frac{k T(-m \zeta_-^2\zeta_+-m\zeta_-\zeta_+^2+\zeta_-^2\gamma+\zeta_-\zeta_+\gamma+\zeta_+^2\gamma)}{mM\zeta_-^3\zeta_+^3}.
\end{equation}
Note that $\kappa $ determines the linewidth $ \Delta f$, and it shows two distinct trends with temperature
\begin{equation} \label{eqn:trend}
\Delta f\sim
\begin{cases}    T^{-7/2} & T<C \\
T^{-1/2} & T>C.
\end{cases}
\end{equation}

\begin{figure} 
\centering
\includegraphics[width=0.45\textwidth]{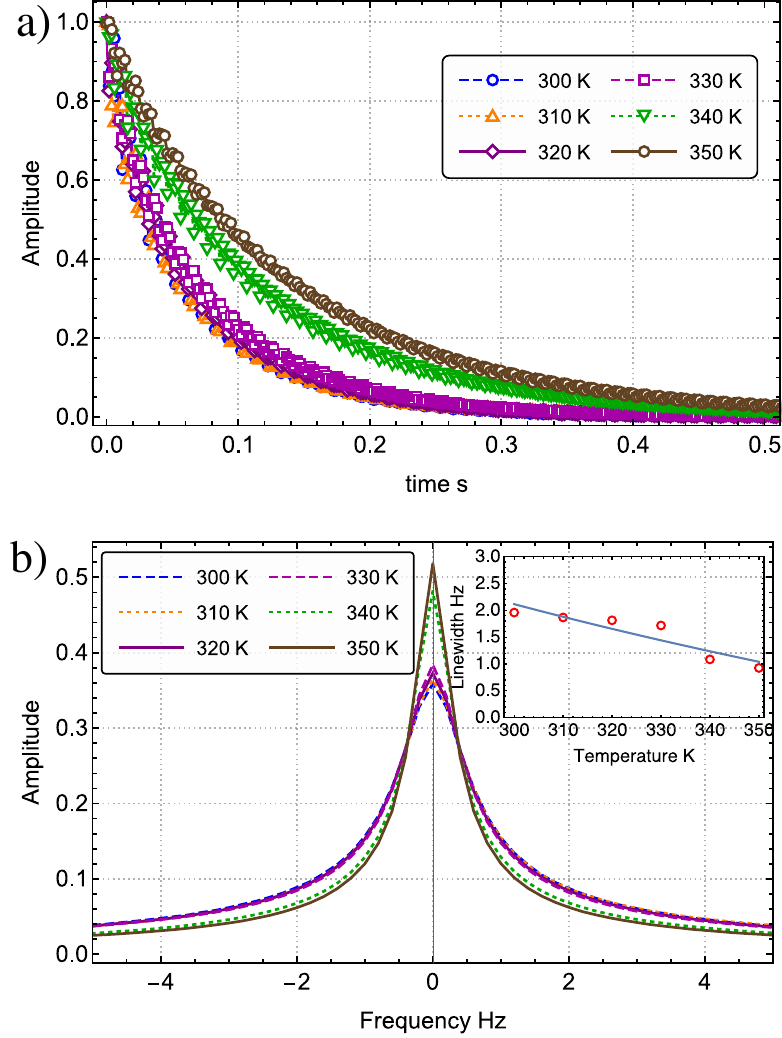}
\caption{a) Shows the decay of NMR echo signal for TMS in gas phase, for variable temperature at $g=0.01$ G/cm. The decay becomes slower at higher temperatures.  Panel b) shows the Fourier transformed data in the frequency domain, illustrating that the NMR spectra  narrows at higher temperatures. The inset indicates the linewidth $\Delta f \propto \kappa(T)$,  extracted by fitting the time domain data to an exponential decay function Eq.~(\ref{eq:nmrsignal}). The blue line shows the best fit using $\kappa(T)=c T^{-1/2}$, where $c$ is a constant.}
\label{fig:exp_gasbulk}
\end{figure}

Figure~\ref{fig:exp_gasbulk} represents the decay of gas phase NMR signal in an echo experiment for a range of temperatures. 
The sample is a sealed tube of liquid tetramethylsilane (TMS), prepared using freeze-pump-thaw method. The sample tube is heated above  $25^\circ$ C to evaporate the TMS. Measurements were performed on a Bruker AV 600 MHz NMR spectrometer equipped with variable temperature and pulsed gradient capabilities.  In the presence of a magnetic-field gradient, the line narrows with increasing temperature, a surprising result that was first observed and explained in~\cite{nature13,PRL15,JCP18}. Data points are fitted to exponential decay functions according to Eq.~(\ref{eq:nmrsignal}) and the resulting linewidths are plotted in the inset of Fig.~\ref{fig:exp_gasbulk}(b). Their temperature trend is in agreement with Eq.(~\ref{eqn:trend}).

{\it The case of liquids.}   To model the viscosity of liquids an empirical equation of the form
\begin{equation}\label{eqn:liqphen}
    \eta=A \exp(B/T)
\end{equation}
where $A$ and $B$ are constants can be substituted into the expression for the nuclear induction decay (c.f. Eq.~\ref{eq:nmrsignal}).  Recalling that $\kappa(T)$ is proportional to $T \zeta^{-3}$, we find:
$$ \kappa(T) \propto T \exp(-3B/T) = T(1-3B/T + \tfrac{9}{2}B^2/T^2 + \dots). $$
The overall temperature dependence of the linewidth
is a line broadening with increasing temperature~\cite{nature13,PRL15,JCP18}.  In any case, these equations are only models for $\eta$.  They do not account for boundaries; yet, it is known that boundaries alter the effective viscosity~\cite{bib:ein1,bib:ein2,bib:ein3,bib:ein4}. Thus, the problem has been reduced to the modeling of an effective viscosity.

\section{\label{sec:exp}MD results}

MD simulations of particle trajectories have been used to model gas diffusion in systems of gas mixtures in complex geometries~\cite{Acosta06,doi:10.1021/jp0477147,doi:10.1021/jp304528d}. In particular, for liquids and non-ideal gases in which transport properties are difficult to describe, MD simulations have proven very successful~\cite{Starov01,Breugem07,Rudyak18,RizkPRL2022}.   

\begin{figure*} [!htb]
    \centering
\includegraphics[width=0.8\textwidth]{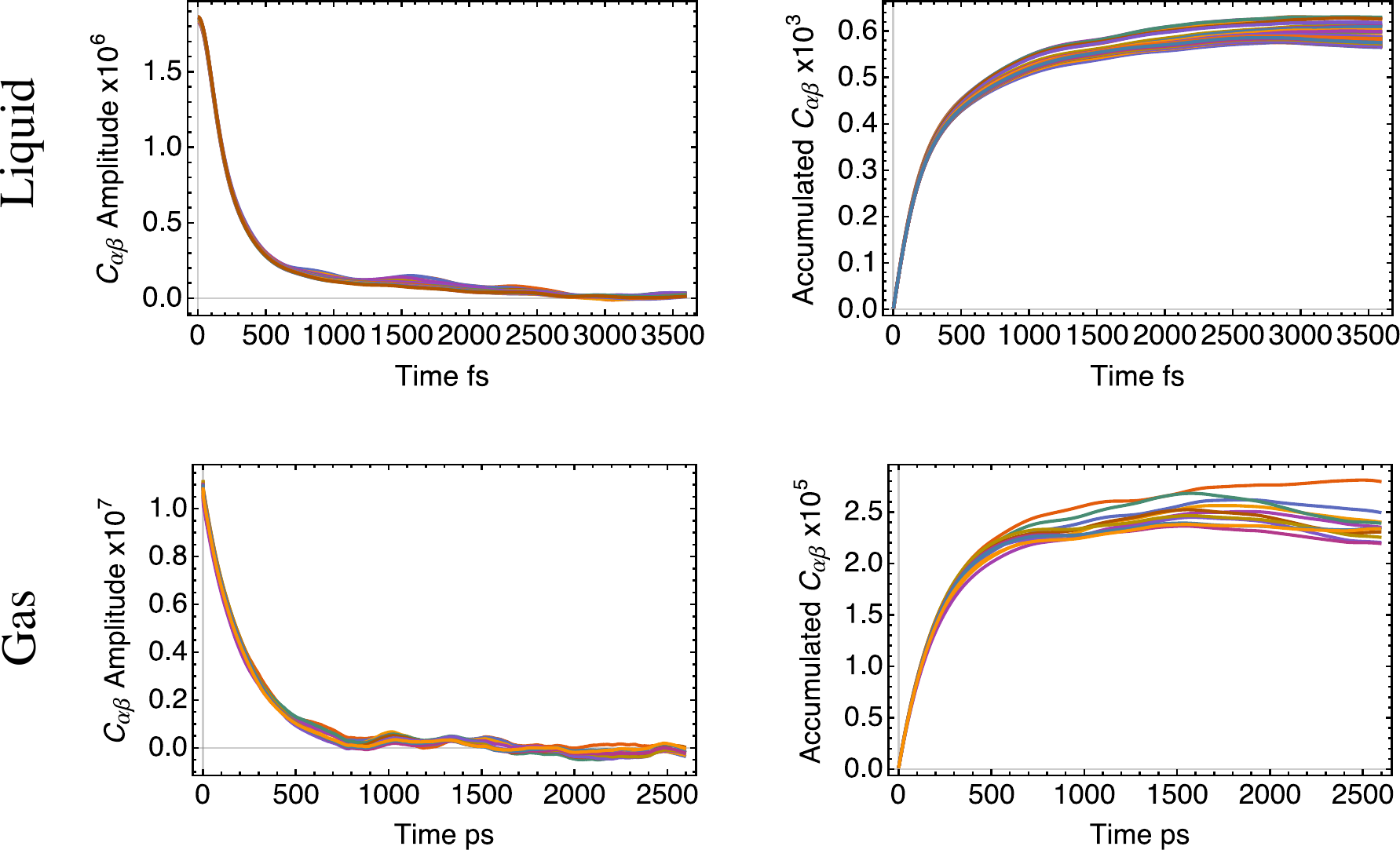}
\caption{ Green-Kubo correlation functions for liquid and gas simulations are evaluated at each temperature for both densities. Here the simulation results at $300 $  is presented (left). These correlation are integrated to the saturation point to evaluate the viscosity coefficient~eq.(~\ref{eq:etaGK}) (right). Note that the time axis is three orders of magnitude larger for gas simulations.}
    \label{fig:corrfun}
\end{figure*}

At each time step, intermolecular forces between nearest neighbors are enforced to recreate realistic particle trajectories~\cite{FRENKEL200263,rapaport_2004}. To obtain such trajectories over a large number of particles we used the open source software ``Large-scale Atomic/Molecular Massively Parallel Simulator'' (LAMMPS)~\cite{LAMMPS}.  LAMMPS results were then used to compute  viscosity coefficients in liquid and gaseous xenon (Xe). We used the Lennard-Jones (LJ) pairing, $U(r) = 4 \epsilon [(\sigma /\ r)^{12} -(\sigma /\ r)^{6}]$, were $\epsilon=1.77$ kJ/mol is the depth of the potential well and $\sigma=4.1~\textup{\AA}$ is the distance where the potential is zero. Simulations were performed for 1,000 Xe atoms in a box with periodic boundary conditions. The simulations ran for a constant number of particles, volume and temperature in the canonical ensemble ($NVT$).   The equilibrium time correlation function approach, a.k.a. Green-Kubo autocorrelation function, was used to derive the viscosity ~\cite{mcq76,rapaport_2004,todd_daivis_2017,Brush62,RizkPRL2022}
\begin{equation}\label{eq:GKintegral}
    \eta= \lim_{t\rightarrow\infty} \eta_{GK}(t)
\end{equation}
with
\begin{equation}\label{eq:etaGK}
\eta_{GK}(t)= \frac{V}{3 k_B T}\int_0^{t} \sum_{\alpha < \beta} C_{\alpha \beta }(\tau) d\tau,
\end{equation}
where $\alpha,\beta \in \{x,y,z\}$, $V$ is the volume and $T$ is the temperature. $C_{\alpha\beta}(\tau)= \left< p_{\alpha\beta}(\tau) p_{\alpha\beta}(0) \right>$ is the autocorrelation function of non-diagonal elements of the pressure tensor, e.g.
\begin{equation*}
p_{xy}(t)=\frac{1}{V}\Bigl\{\sum_j m_j v_{jx}(t) v_{jy}(t)+\frac{1}{2} \sum_{i\neq j} r_{ijx}(t) f_{ijy}(t)\Bigr\}.
\end{equation*}
Here $f_{ijy}$ represents the $y$-component of the force between two particles $i$ and $j$. The first term on the right hand side, is the kinetic contribution to the pressure tensor while the second term indicates the potential contribution. Other components $p_{\alpha\beta}$ of the pressure tensor are defined analogously.

Simulations were performed for 21 different temperature values in the range 200-400 K, using a high density ($\rho=N/V$) of particles compatible with liquid phase, and a low density for gas phase. Particle trajectories were evolved for $10^6$ femtoseconds to equilibrate. We evaluated $C_{\alpha \beta}$ correlation functions 60 times in each simulation, recording them well beyond their saturation point (see Fig.~\ref{fig:corrfun}). They were then used to evaluate the  Green-Kubo integral Eq.~\eqref{eq:GKintegral}. For this approach to work, it is critical that the correlation functions decay with similar rates so the resulting integrals converge to an average value.

\begin{figure}[!tbh]
\centering
\includegraphics[width=0.45\textwidth]{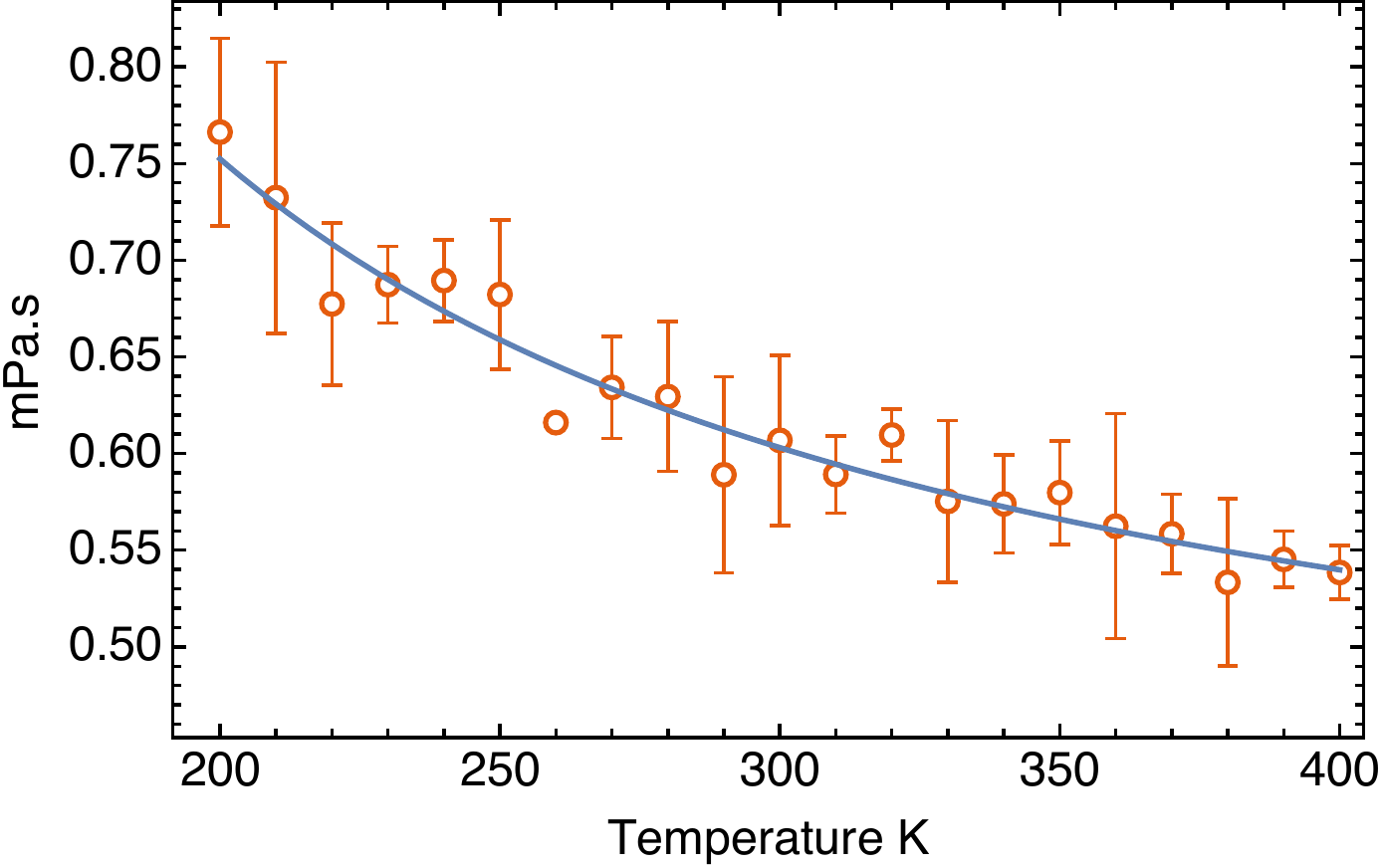}
\caption{Viscosity in liquids drops as the temperature increases. Error bars indicate the standard deviation for the set of 60 accumulated correlation functions. The solid line is a fit to Eq.~(\ref{eqn:liqphen}) with $A=0.38\pm0.01$ and $B=135\pm7$.}
\label{fig:liqvisc}
\end{figure}

The shear viscosity coefficient for the liquid state was found to decrease with temperature (see Fig.~\ref{fig:liqvisc}). This behavior is expected and well understood for liquids, and is in agreement with the line broadening observed in the liquid state NMR experiments~\cite{nature13,PRL15}.  The verification of simulated viscosity coefficients against experimental values is complicated by the fact that Xe is not in the liquid form above 270 K whereas below that temperature, measurements report viscosity vs temperature but across different densities~\cite{hanley1974viscosity,grisnik1998measurement}. Nevertheless, a matching data point exists: at 220 K our simulated viscosity coefficient lies within $5\%$ of the experimentally measured value~\cite{hanley1974viscosity}. Our simulation results are similar to those from Ref.~\cite{rowley1997diffusion} where the authors performed a general study of LJ potential dynamics vs $\rho$ and $T$ and found that the resulting viscosity coefficients are within $6\%$ of experimental values. The decay of $\eta(T)$ with $T$ was found to be exponential and was fitted to Eq.~(\ref{eqn:liqphen}).

\begin{figure} [!bth]
\centering
\includegraphics[width=0.45\textwidth]{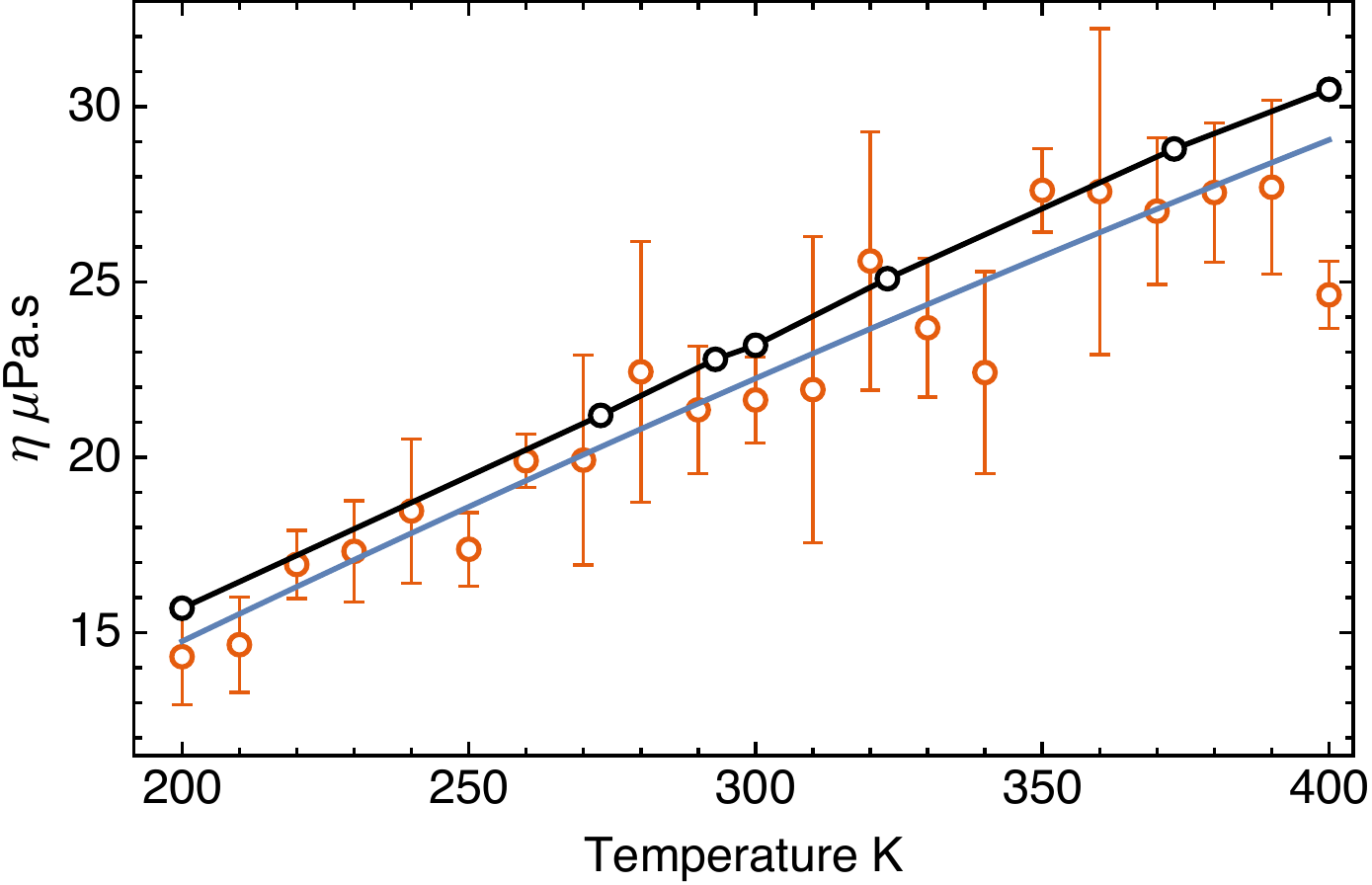}
\caption{Viscosity in gas molecules grows as the temperature increases. Red circles indicate results of MD simulation and black circles are experimental data. Error bars are the standard deviation for the set of 60 accumulated correlation functions. The solid line shows a fit to Eq.~(\ref{eq:gasviscform}) with $C=258\pm69$.}
\label{fig:gasvisc}
\end{figure}

The gas phase simulations were  performed by first equilibrating the system over a period of $10^6$ picoseconds. Comparing to the liquid case, the correlation functions were found to decay slower as the frequency of particle collision is smaller (see Fig.~\ref{fig:corrfun}). The simulated viscosity values were found to be two orders of magnitude smaller as well.  Moreover, viscosity coefficients in gases show the opposite trend to those of liquids, i.e. shear viscosity increases with temperature. A comparison to the available experimental data~\cite{crchandbook,mountain2007molecular} shows that the simulated viscosity coefficients  overlap with them. The calculated viscosity coefficients are fitted to Eq.~(\ref{eq:gasviscform}). We find that the Sutherland's constant ($C=258 \pm 69$) is in agreement with the literature value for Xe gas ($C=252$)~\cite{crchandbook}.

\begin{figure} [!tbh]
\centering
\includegraphics[width=0.45\textwidth]{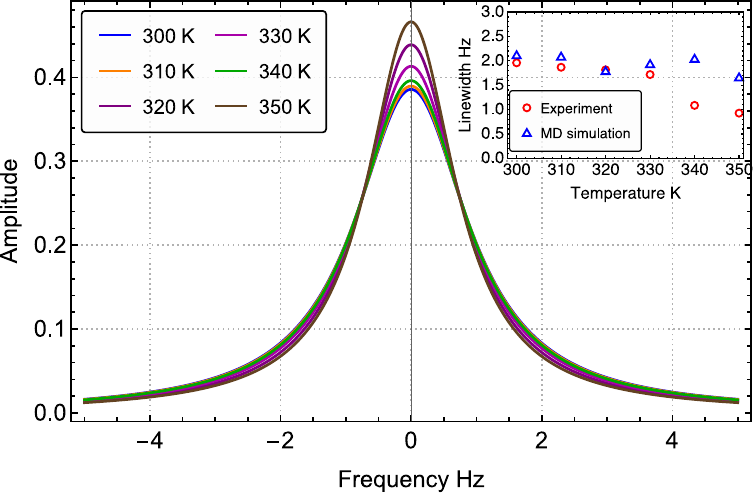}
\caption{Dependence of nuclear induction lineshape for Xe gas on temperature. MD-derived viscosity coefficients were used together with Eq.~(\ref{eq:kappa}). The inset compares experimental linewidths with simulated ones.}
\label{fig:gasline}
\end{figure}

Our MD results confirmed that as the temperature of gases increases, the higher frequency of molecular collisions results in greater resistance and larger viscosities. Viscosity results are in agreement with previous studies and experimental results, as expected. On the other hand, what is new here is our ability to predict the correct lineshape for the nuclear induction signal from MD results, and this lineshape is predicted based on the effective viscosity. Equation~(\ref{eq:kappa}) suggests that the decay rate of the nuclear induction signal is inversely proportional to the viscosity coefficient $\eta$ and as a result, the linewidth narrows at higher temperatures. A depiction of such behavior for the experiment with $g=0.01$ G/cm, is shown in Fig.~\ref{fig:gasline} where the simulated viscosity coefficients were used to predict the lineshape in gases.    The resulting linewidths are very close to the experimental values of Fig.~\ref{fig:exp_gasbulk}.  A side-by-side comparison of experiments and theory is shown in the Inset of Fig.~\ref{fig:gasline}.

\section{Conclusion}

We have shown the feasibility of using MD simulations to model transport coefficients in liquids and gases; these transport coefficients are then used together with a GLE model to obtain the nuclear induction lineshape. In particular, our MD results correctly indicate that gas phase spectra exhibit narrower lines at higher temperatures. (This is in contrast to the conventional formalism~\cite{HerzogHahn,PFITSCH1999,slichterbook}, which incorrectly predicts the opposite trend.) This work opens the door for the prediction of lineshapes in more complex geometries and boundary conditions.  Our work differs from previous studies of diffusion~\cite{grebenkov2020paradigm} in that memory effects can be explicitly modeled using our method.  Future directions may include explicit modeling of boundaries and gas types using appropriate memory kernels, and ultimately, solving the inverse problem of computing molecular and pore parameters from experimentally-measured nuclear induction decays.

\section{Acknowledgment}
This work used computational and storage services associated with the Hoffman2 Shared Cluster provided by UCLA Institute for Digital Research and Education’s Research Technology Group.  L.-S. B. acknowledges funding from NSF award CHE-2002313.

\bibliographystyle{naturemag}
\bibliography{MDlw.bib}



\end{document}